\documentclass[twocolumn]{aastex631}

\newcommand{\msun}{\(M_\odot\) }
\newcommand{\rsun}{\(R_\odot\) }
\newcommand{\chandra}{\textit{Chandra }}
\newcommand{\XMM}{\textit{XMM-Newton }}
\newcommand{\swift}{\textit{Swift }}

\shorttitle{Stellar Wind of WR Star in IC\,10\,X-1}
\shortauthors{Bhattacharya et al.}
\graphicspath{{./}{}}

\begin{document}

\title{Probing the Stellar Wind of the Wolf-Rayet Star in IC\,10\,X-1}

\correspondingauthor{Sayantan Bhattacharya}
\email{sayantan34@gmail.com}

\author[0000-0001-8572-8241]{Sayantan Bhattacharya}
\affiliation{Department Of Physics \& Applied Physics, University Of Massachusetts, Lowell, MA, 01854}
\affiliation{Lowell Center for Space Science \& Technology, University Of Massachusetts, Lowell, MA, 01854}

\author[0000-0002-8427-0766]{Silas G. T. Laycock}
\affiliation{Department Of Physics \& Applied Physics, University Of Massachusetts, Lowell, MA, 01854}
\affiliation{Lowell Center for Space Science \& Technology, University Of Massachusetts, Lowell, MA, 01854}

\author[0000-0002-1115-6559]{Andr\'e-Nicolas Chen\'e}
\affiliation{Gemini Observatory/NSF’s NOIRLab, 670 N. A‘ohoku Place, Hilo, Hawai‘i, 96720, USA}

\author[0000-0002-4955-0471]{Breanna A. Binder}
\affiliation{Department of Physics \& Astronomy, California State Polytechnic University, Pomona, CA 91768}

\author[0000-0002-7652-2206]{Dimitris M. Christodoulou}
\affiliation{Lowell Center for Space Science \& Technology, University Of Massachusetts, Lowell, MA, 01854}

 \author[0000-0001-6068-203X]{Ankur Roy}
 \affiliation{Department Of Physics \& Applied Physics, University Of Massachusetts, Lowell, MA, 01854}
 \affiliation{Lowell Center for Space Science \& Technology, University Of Massachusetts, Lowell, MA, 01854}

 \author[0000-0002-3562-9699]{Nicholas M. Sorabella}
 \affiliation{Department Of Physics \& Applied Physics, University Of Massachusetts, Lowell, MA, 01854}
 \affiliation{Lowell Center for Space Science \& Technology, University Of Massachusetts, Lowell, MA, 01854}

 \author[0000-0003-0267-8432]{Rigel C. Cappallo}
 \affiliation{Massachusetts Institute Of Technology, Haystack Observatory, Westford, MA, 01886, USA}



\begin{abstract}

    IC 10 X-1 is an eclipsing high mass X-ray binary (HMXB) containing a stellar-mass black hole (BH) and a Wolf-Rayet (WR) donor star with an orbital period of $P = 34.9$ hr. This binary belongs to a group of systems that can be the progenitors of gravitational wave sources, hence understanding the dynamics of systems such as IC\,10\,X-1 is of paramount importance. The prominent He II 4686 emission line (previously used in mass estimates of the BH) is out of phase with the X-ray eclipse, suggesting that this line originates somewhere in the ionized wind of the WR star or in the accretion disk. 
    We obtained 52 spectra from the \textit{GEMINI}/GMOS archive, observed between 2001 and 2019. We analyzed the spectra both individually, and after binning them by orbital phase to improve the signal-to-noise ratio.  The RV curve from the stacked data is similar to historical results, indicating the overall parameters of the binary have remained constant. However, the He II line profile shows a correlation with the X-ray hardness-ratio values, also,  we report a pronounced skewness of the line-profile, and the skewness varies with orbital phase. These results support a paradigm wherein the \ion{He}{ii} line tracks structures in the stellar wind that are produced by interactions with the BH's ionizing radiation and the accretion flow. We compare the observable signatures of two alternative hypotheses proposed in the literature: wind irradiation plus shadowing, and accretion disk hotspot; and we explore how the line-profile variations fit into each of these models.

\end{abstract}



\section{Introduction}

Black hole binary (BHB) systems are one of the endpoints for massive stellar evolution, and the powerhouse of a certain class of gamma-ray bursts \citep{israelian1999, brown2000grb,orosz2001}. 
 Most of the currently known BHBs are found in low mass X-ray binaries \citep[LMXBs,][]{remillard2006x}, due to the longer lifetimes of the low mass donors. Although rare, BHBs containing a massive Wolf-Rayet (WR) star are crucial to understanding the evolution of massive stars. These systems have very small orbital separations (and periods), which suggests that they have gone through a phase of significant shrinking from their initial separations, i.e., from $\sim$100-1000\rsun down to $<$10\rsun\citep{psaltis, tanaka_lewin}. This shrinking likely happened during their common-envelope phase, before the formation of the compact object \citep{whelan1973binaries, iben1984supernovae, tauris2017formation, igoshev2021combined}. After core collapse of the WR star, if the system happens to remain bound, and the orbit does not widen significantly, the binary system can result in a binary with two revolving compact objects. If the separation becomes small enough, these stars will eventually spiral in and merge, generating gravitational waves that can be detected from Earth. Hence, BH-HMXBs are of great importance as progenitors of gravitational wave sources, providing observational constraints on the predicted rate of current LIGO/VIRGO detections \citep{belczynski2013cyg}.

This work presents our investigation on the extragalactic BHB IC\,10\,X-1, a high mass X-ray binary (HMXB) consisting of an accreting BH and a WR-star donor. In HMXBs, X-rays are commonly produced, as the wind from the donor star is accreted by the companion compact object. The interaction between the X-rays from the BH and the stellar wind can be seen in the spectral characteristics of the system \citep{remillard2006x}. 
IC\,10\,X-1 was discovered as the most luminous X-ray binary in the starburst galaxy IC\,10 \citep{brandt1997rosat}. Observations with {\it Chandra}, \textit{GEMINI}, and {\it Hubble} have confirmed the X-ray source's association with the WN3-type WR star [MAC92]17A \citep{bauer2004chandra,clark2004wolf}. Subsequently, \swift and \chandra observations were used to determine the system's orbital period as 34.4$\pm$0.83 hr \citep{prestwich2007orbital}, and the compact object was proposed to be a black hole. Using the radial velocity (RV) curve obtained from the \ion{He}{ii}\,$\lambda$4686 emission line, \citet{silverman2008} determined the masses of the WR star and the compact object to be $\sim$35\msun and $>23$-$24 M_\odot$, respectively, albeit with large uncertainties. \citet{laycock2015} performed a study using a decade's worth of \chandra and \XMM data, which revealed that the system has a 5-hour-long (asymmetric) eclipse, and significantly improved the orbital ephemeris  of the system ($P = 34.8419\pm0.0002$ hr , $T_0$= 54040.87 MJD to place mid-eclipse at $\Phi=0.5$). These datasets led to the surprising discovery that the mid-point of the X-ray eclipse and the zero-velocity of the RV curve did not match \citep{laycock2015}. In fact, the eclipse midpoint coincides with the maximum blueshift of the line.  A similar behavior had been seen in another WR+BH HMXB Cygnus X-3, in its low hard state only \citep{van1996wolf, koljonen2017gemini}, and has since been observed in NGC 300 X-1 \citep{Carpano2019, binder2021wolf}. This puts the determined BH masses of these systems into question, as the premise of using the \ion{He}{ii}\,$\lambda$4686 line as a tracer of the WR star's motion may not be correct.

The stellar wind from IC 10 X-1 has a highly ionized part due to irradiation by X-rays emanating from the compact object, and a relatively cooler shadowed region on the side of the WR star opposite the BH. The degree of ionization is controlled by the parameter $\xi = L_X/(nd^2)$ \citep[where $n$ is the material density at distance $d$ from the source;][]{tarter1969ion}. For the IC 10 X-1 system \cite{barnard2014energy} compute $\xi \approx 10^3$~erg~cm~s$^{-1}$, a level at which neutral hydrogen cannot exist. 

The \ion{He}{ii}\,$\lambda$4686 line may in that case have its origin in the shadowed section of the stellar wind. The line centroid and the line profile may therefore exhibit a substantial contribution from the wind velocity, and the observed RV curve will be modulated by the binary orbit. Hence, the RV semi-amplitude will provide the WR wind velocity (at the depth in the wind where the emission line originates) \citep{van1996wolf,laycock2015revisiting,bhattacharya2021mass,bhattacharya2022stellar}.

Bulk ionization of the wind can in turn interrupt the wind acceleration mechanism (WR winds being dominated by line-driving) and lead to the formation of a photoionized ``wake.'' 
The slow moving wake can become gravitationally captured by the compact object, enhancing the accretion rate and forming  a persistent accretion disk.  \ion{He}{ii}\,$\lambda$4686 line emission may originate in the outer edge of the accretion disk, at or near a "hotspot" where the material from the donor star impacts the disk \citep{el2019formation,el2019wind, binder2021wolf}. If part of the \ion{He}{ii}\,$\lambda$4686 originates near the accretion disk/hotspot, the resulting radial velocity(RV) component will trace the orbital motion at that location, having a different velocity amplitude and phase shift. 

We aim to search for evidence of the above effects using a number of archival optical spectra of IC\,10\,X-1. 
The outline of this paper is as follows. In section 2, we describe the archival data used in this study and the data reduction procedures. In section 3, we present the RV measurements from our study and an analysis of the \ion{He}{ii}\,$\lambda$4686 emission line profile. In section 4, we discuss the trends in the \ion{He}{ii} line's normalized skewness and their coupling to the WR stellar wind. Finally, in section 5, we conclude with a summary of our findings.
\begin{table*}
    \centering
    
    \begin{tabular}{l l c c c c c}
        \hline \hline
        Program ID   & Start Date & End Date & No. Of  & Mask Name & Object & PI \\ 
                 &   &   &  Spectra & & ID & Name\\
        \hline
        GN-2001B-Q-23 & 2001-12-22 & 2002-01-17 & 5 & GN2001B-Q-23-1 & 15916 & Crowther, P. A $_{[1], [2], [3]}$\\ \\
        GN-2004B-Q-12 & 2004-07-16 & 2004-08-13 & 9 & GN2004BQ012-01 & 66 & Massey, P. M.\\ \\
        GN-2010B-Q-58 & 2010-09-02 & 2010-09-07 & 6 & GN2010BQ058-01 & 16 & Laycock, S. G. T. $_{[4], [5]}$\\ \\
        GN-2017B-Q-20 & 2017-11-10 & 2017-11-15 & 7 & GN2017BQ020-03 & 37 & Massey, P. M.\\ \\
        GN-2018B-Q-127 & 2018-12-13 & 2019-01-04 & 10 & GN2017BQ046-01 & 20 & Laycock, S. G. T.\\ \\
        GN-2018B-Q-127&2019-07-02 &2019-07-04 & 15 & GN2018BQ127-03 & 220 & Laycock, S. G. T.\\ \\
         \hline \hline
         
    \end{tabular}
    \caption{\textit{GEMINI}/GMOS available archival data. Previous publications (as listed in \textit{GEMINI} archive) using these : [1] \citet{crowther2003gemini}, [2] \citet{clark2004wolf}, [3] \citet{prestwich2007orbital}, [4] \citet{laycock2014transient}, [5] \citet{laycock2017x}}
    \label{tab:gmos_data}
\end{table*} 
\begin{figure}
    \centering
    \includegraphics[width = 0.45\textwidth]{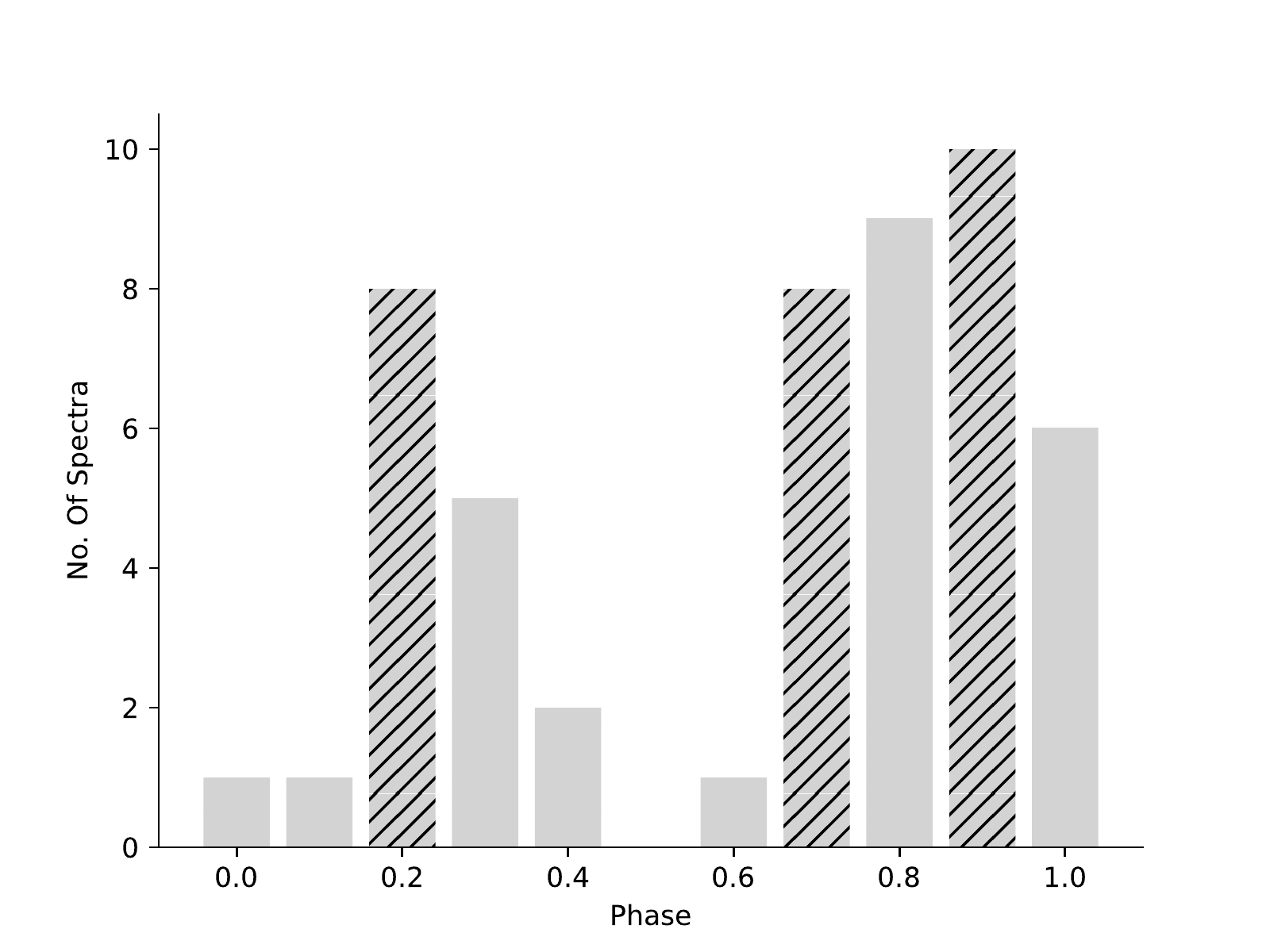}
    \caption{IC\,10\,X-1 phasewise spectra availability in the GMOS archive}
    \label{fig:ic10_hist}
\end{figure}
\section{Observations \& Data Analysis}
We have used the rich dataset from the \textit{GEMINI/GMOS} archive, which consists of 52 spectra of IC\,10\,X-1 (also known as [MAC 92]17A), spanning the years 2001 to 2019. IC\,10\,X-1 was one of the many targets observed in multi-object spectroscopy (MOS) mode using custom slit-masks and the B600 grating. There are also some longslit spectra available in the archive, acquired using the R150 grating. These spectra were not used in the present work as a precise wavelength calibration could not be performed due to missing arc data, and the spectral resolution was in any case too coarse. All of the MOS spectra were obtained using the B600 grating and they belong to the five \textit{GEMINI} queue mode observing programs listed in Table\,\ref{tab:gmos_data}. We reprocessed the data using the \textit{gemini package} in {\sc iraf}\footnote{{\sc iraf} was distributed by the National Optical Astronomy Observatory, which was managed by the Association of Universities for Research in Astronomy (AURA) under a cooperative agreement with the National Science Foundation.} and following the commands dictated in \textit{gmosexample}, under the MOS mode.

We followed the usual steps (see, e.g., \textit{gmosexample}) for overscan, bias, and flatfield correction. The wavelength calibration was done using the spectra for the internal GMOS CuAr lamp. The spectra were corrected for quantum efficiency using the wavelength calibrated flatfield frame for reference. The cosmic rays were removed using the L.A. Cosmic algorithm from {\sc iraf}. Instead of using \textit{gssubstract} for sky subtraction, we fitted a narrow region of the trace, thus avoiding contamination from a nearby star. This procedure was carried out as part of the \textit{gsextract} command, which extracts the 1-D spectra from the 2-D images.

These are all GMOS archive data collected for different science programs, hence the exposure time for IC\,10\,X-1 was not always optimal for our purposes and the signal-to-noise ratio (SNR) was at times quite low, even in the centers of the spectra, where the \ion{He}{ii}\,$\lambda$4686 line is present. Having 52 usable spectra we applied the above ephemeris to calculate the phase at which each spectral exposure was observed, and created a histogram of phase coverage, as shown in Figure \ref{fig:ic10_hist}. We then created stacked  spectra with increased signal-to-noise ratio by combining the spectra obtained in each orbital phase-bin of width 0.1. Coverage was not uniform: six phase bins contain 5 or more spectra (one has 10), 3 phase-bins have a single exposure only, and one phase-bin lacks coverage altogether. Regrettably, the missing phase is the one centered on $\Phi=0.5$, i.e. the anticipated maximum blueshift at mid-eclipse. Future observing efforts are encouraged to obtain spectra at this phase, and to increase coverage of the other bins to achieve better uniformity across phases. 

In order to explore whether changes in the ionizing X-ray flux were correlated with optical line profile changes,  X-ray hardness ratios (HRs) were calculated from archival \textit{Chandra} data. The raw data were reduced using standard procedures in CIAO 4.13, i.e., cleaned and reprocessed event files were created using the \texttt{chandra repro} routine, and they were barycenter-corrected using the \texttt{axbary}; then, the background light curves were created in the broad (0.5 - 7keV), hard (3 - 7 keV), and soft (0.5 - 1.2 KeV) energy bands using the \texttt{dmextract} command with a 7 ks time bin. The source counts were extracted from a 20$^{\prime\prime}$ circular region, and the background was calculated using an annular region with 30$^{\prime\prime}$ and 50$^{\prime\prime}$ as inner and outer radii, respectively. The hardness ratios (defined by (H-S)/(H+S), where H is the hard net counts and S is the soft net counts) were calculated using the Bayesian estimation software BEHR \citep{park2006bayesian}.
\begin{table}
    \centering
    \caption{\textit{Chandra} data used to determine hardness-ratio measurements}
    \begin{tabular}{c c c c}
        \hline \hline
        Observation & Date & MJD Date & Exposure   \\ 
           ID.      &   &   &  (ks)\\
        \hline
        3953 & 2003-03-12 & 52710.7 & 29.2 \\
        7082 & 2006-11-02 & 54041.8 & 42.6 \\
        8458 & 2006-11-04 & 54044.2 & 43.6 \\
        11080 & 2009-11-05 & 55140.7 & 14.8 \\
        11081 & 2009-12-25 & 55190.2 & 13.7 \\
        11082 & 2010-02-11 & 55238.5 & 14.9 \\
        11083 & 2010-04-04 & 55290.6 & 14.7 \\
        11084 & 2010-05-21 & 55337.8 & 14.6 \\
        11085 & 2010-07-20 & 55397.5 & 14.9 \\
        11086 & 2010-09-05 & 55444.6 & 14.9 \\
         \hline \hline
    \end{tabular}
    
    \label{tab:cxo_data}
\end{table} 

\section{Analysing the \ion{He}{ii} $\lambda$4686 emission line}

\begin{figure}
    \centering
    \includegraphics[width = 0.5\textwidth]{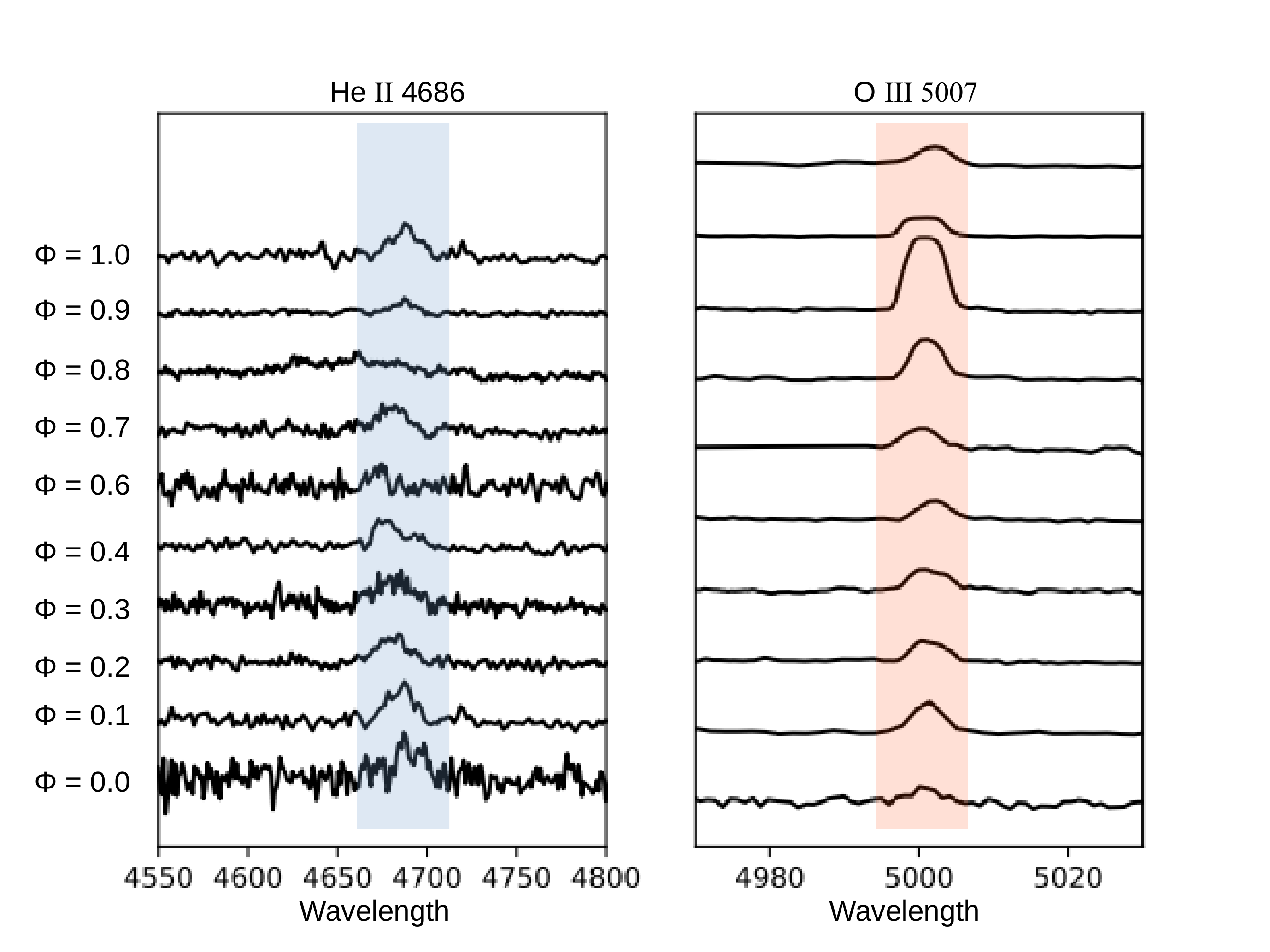}
    \caption{The phase-wise stacked spectra of IC~10 X-1, with a relative flux offset. The He II (left) and O III (right) emission lines are shown. The O III line traces the bulk motion of the IC\,10 galaxy itself, and it was used to correct for the bulk motion of IC 10 in the calculations of the He~II RV.}
    \label{fig:ic10_spec}
\end{figure}

The \ion{He}{ii} $\lambda$4686 emission line is the only detectable stellar spectral feature in the GMOS spectra that is emitted by IC\,10\,X-1 (and not Galactic), and that is strong enough to be detected (see Figure~\ref{fig:ic10_spec}). Several prominent nebular emission lines are visible, for example, the nearby O III ($\lambda$5007) which occurs blueshifted in all of the spectra, corresponding to the relative velocity of IC 10 with respect to the solar system. H$\alpha$ also appears in emission at the IC 10 velocity, due to the H II region in which X1 appears embedded.

Using the \textit{splot} command in {\sc iraf} we measured the centroid of the He II line and the O III line in each spectrum.  We also measured the equivalent width (EW), full-width at half maximum (FWHM), and skewness of the He II line. We discuss these parameters below. 

\subsection{Radial Velocity Curve}

Phasewise stacked spectra were used for creating the RV curve for IC\,10\,X-1. The RV values were calculated by measuring the shift of the line's centroid with respect to the rest-frame value ($\lambda_0$). The centroid of the \ion{O}{iii}\,$\lambda$5007 narrow emission line was assumed to track the velocity of the IC\,10 galaxy, and it was used as the rest velocity of the \ion{He}{ii}\,$\lambda$4686 line. 

\begin{figure}
    \centering
    \includegraphics[trim = {0 0 0 1.28cm}, clip, width= 0.5\textwidth]{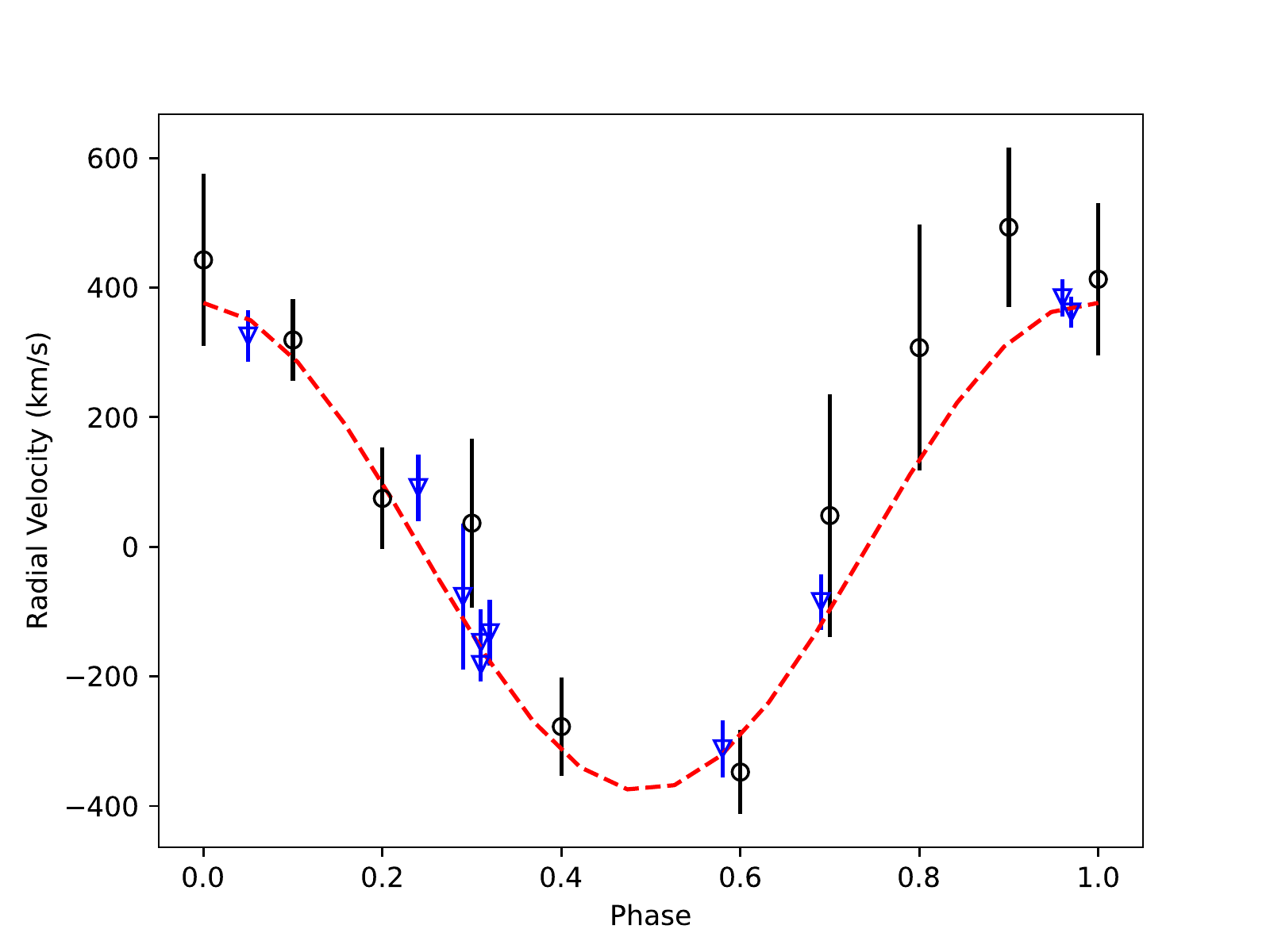}
    \caption{Radial velocity curve of IC\,10\,X-1. The black (round) and blue (triangles) markers show GMOS stacked data and \citet{silverman2008} Keck data, respectively.}
    \label{fig:rv_stacked}
\end{figure}

The resulting \textit{GEMINI} RV values are plotted versus orbital phase in Figure \ref{fig:rv_stacked}, to which we have added the literature RV values of \cite{silverman2008} (that were derived from an independent Keck dataset).  The RV data were fitted with a sine function, where the velocity offset was set to zero because the velocity is already corrected for the IC\,10 galaxy's motion. However, it is not clear whether the space velocity of X-1 is really zero with respect to IC 10.

The best fit to the combined (\textit{GEMINI} + \textit{Keck}) RV points yields a velocity semi-amplitude of 376.7 $\pm$12.4\,km\,s$^{-1}$ which is consistent with  the previous results of \citet{silverman2008}. Both datasets show the same 0.25 phase-offset with respect to the X-ray eclipse ephemeris as reported by \citet{laycock2015}. 

\subsection{Equivalent Width of \ion{He}{ii} $\lambda$4686 Emission Line}

\begin{figure}
    \centering
    \includegraphics[trim = {3cm 6cm 4cm 4.5cm}, width= 0.35\textwidth, height = 0.5\textwidth]{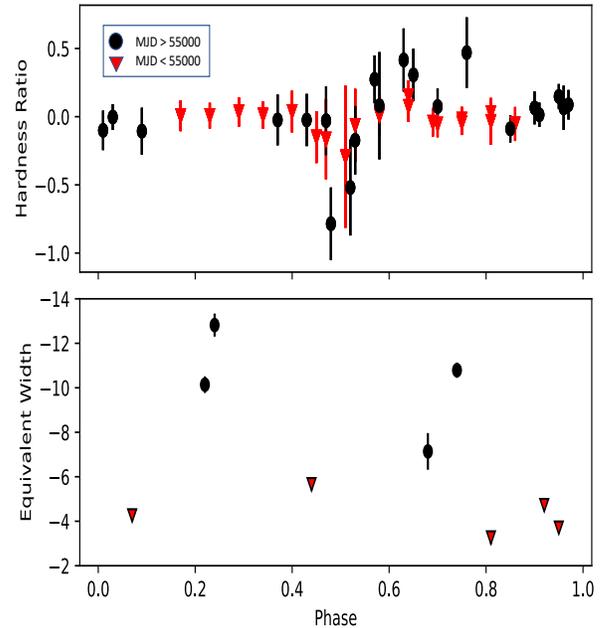}
    \caption{HR and He~II EW variation with phase $\phi$. The system transitions to a softer state going into the eclipse, whereas there is no such evident phase dependence for the He~II EW. The X-ray emission is harder coming out of the eclipse ($\phi$ = 0.6-0.8) for the 2009 (MJD 55140.7) data than in previous observations; the EWs also increased at that time.}
    \label{fig:hr_eq_phase}
\end{figure}

We measured the EW[He II] in order to look for variability over time and with phase (see Figure \ref{fig:hr_eq_phase}). Such changes would indicate variations in the amount of emitting material, which in turn might be controlled by changes in ionizing flux  in response to the BH accretion state. Figure \ref{fig:hr_eq_phase} shows both EW[He II] and X-ray HR across binary phase.  Generally, the HR follows a pattern where it is steady out of eclipse and even during the initial part of ingress, declines sharply at mid-eclipse and then gets harder throughout egress, after which it declines back to its `steady' level. Here we have selectively used optical data from programs GN-2001B-Q-23 (taken in 2001) and GN-2010B-Q-58 (taken in 2010), and their contemporaneous X-ray data. These time ranges were chosen because the HR (averaged across whole observations) of X-1 had shown a significant increase relative to its typical constant values on 5 Nov 2009 (MJD 55140.7) from \chandra observations, and the HR values kept falling toward normal values, as they were last observed on 5 Sep 2010 (the next observation of this system was in 2012 using \XMM), as shown in Figure 1 of \citet{laycock2015}. In Figure \ref{fig:hr_eq_phase}, we see that the 2010 HR profile (black points) follows the usual pattern except that it is substantially harder during/after egress. The EW[He II] values  for the contemporaneous 2010 GMOS observations are also  significantly higher than in 2001. The average EW[He II] goes from $-4 \AA$ in 2001, to $-10 \AA$ in 2010. Due to the sparse data it is not possible to definitively say whether this change is in response to changes in the X-rays or a periodic variation of EW[He II] with phase. In the range of $\Phi=0.7$-0.8, we do see differing EW values between 2001 and 2010, suggesting a correlation between the HR and EW. New optical spectral coverage at all phases would be useful to test this hypothesis.  

\subsection{Skewness of \ion{He}{ii} line and The Stellar wind}

Skewness is a measure of deviation from a symmetric profile. Measurement of the skewness of a particular spectral line can reveal interesting physics concerning the environment where the line originates, for example the velocity distribution of the ions producing the emission. We have calculated the normalized skewness of the \ion{He}{ii}\,$\lambda$4686 line from our phase-stacked spectra.  Skewness is the 3$^{\rm rd}$ moment of the distribution, where the $n^{\rm th}$ moment is defined as
\begin{equation}
   u(n) = \frac{\int(\lambda - \lambda_0)^n F_\lambda d\lambda}{\int F_\lambda d\lambda}\, ,
\end{equation}
where
\begin{equation}
   \lambda_0=\frac{\int \lambda F_\lambda d\lambda}{\int F_\lambda d\lambda}.
\end{equation}
The skewness values for \ion{He}{ii}\,4686 line were normalized by the second moment $u(2)$ and the values that we use here are given by
\begin{equation}
{\rm Normalized\ Skewness} = u(3)/[u(2)]^{3/2}.
\label{norm_skew}
\end{equation}

\begin{figure}
    \centering
    \includegraphics[width=0.5\textwidth]{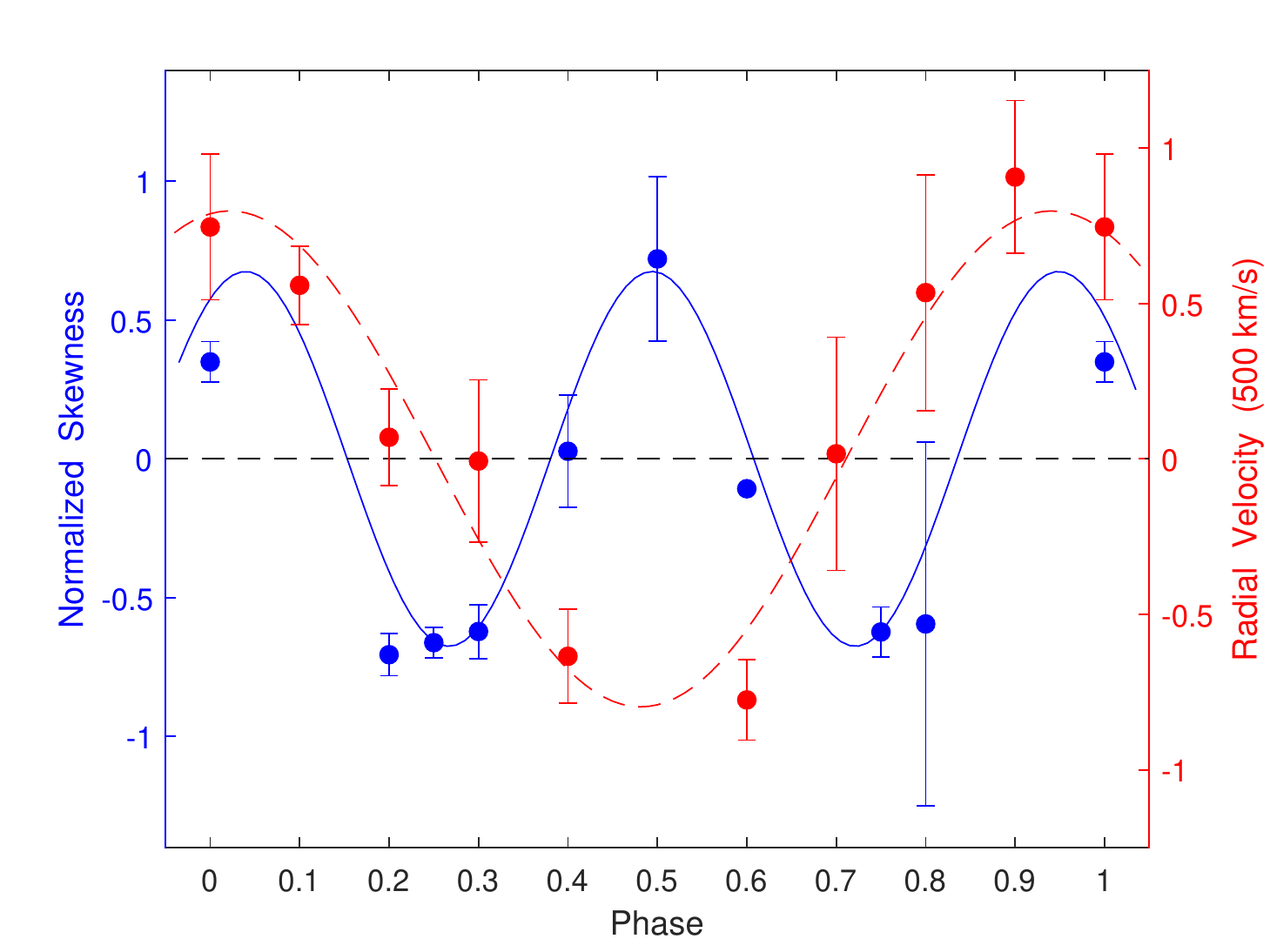}
    \caption{Variation of skewness values for the \ion{He}{ii} line with phase in blue and the RV variation in red. Skewness compltetes 2 cycles within each RV cycle. Anticorrelation is evident in the phase interval [0.25, 0.75].}
    \label{fig:skew_phase}
\end{figure}

In addition to the original phase combined spectra (with 0.1 bins), we have also used another set of spectra where we binned the data in 0.25 phase bins ($\phi_0$=0.875-0.125, $\phi_{0.25}$=0.125-0.375, $\phi_{0.5}$=0.375-0.625 and $\phi_{0.75}$=0.625-0.875). The normalized skewness values from this combined set are shown plotted against orbital phase in Figure \ref{fig:skew_phase}. The results are displayed along with RV measurements from this work for comparison purposes (the older Keck data were not included here, hence the red dash curve is different from the RV fit in fig.\ref{fig:rv_stacked}). The RV values are normalized to 500 km~s$^{-1}$ and the skewness values are normalized as shown in equation~(\ref{norm_skew}). The errors in skewness were calculated using the Poison ($\sqrt{N}$) errors in counts $N$ and equation~(\ref{norm_skew}) to obtain both upper and lower bounds. We see that the line profile alternates between negative and positive skewness. In this context, negative values mean the line is skewed to the shorter wavelength (blue) side and positive values mean that the skewness lies on the long wavelength (red) side. Both sets of normalized values were fitted with periodic cosine functions. Skewness has an amplitude of $S_0= 0.68\pm0.07$, a period of $P_S=0.45\pm0.01$ in terms of phase, and a horizontal shift of $\alpha=-31\pm13$ degrees. The corresponding best-fit RV values are $V_0=0.80\pm0.07$, $P_V=0.92\pm0.06$, and $\alpha=-9\pm13$ degrees, respectively. The two sinusoidal curves are out of phase by about $20\pm18$ degrees, and the period ratio is $P_S/P_V=0.49\pm0.03$. Because of these values, skewness and RV are strongly correlated in quadrants 4 and 1 (phases in [0.75, 1.25]), and clearly anticorrelated in quadrants 2 and 3 of the phase (where the BH is eclipsed by the WR star, at phases centered around 0.5).

\section{Discussion}

The IC 10 X-1 binary system showed possible correlations in X-ray HR values and the EW values of the \ion{He}{ii} $\lambda$4686 line (Figure \ref{fig:hr_eq_phase}). The variation in hardness ratio can either suggest the presence of an obscuring column in the line of sight, or changes in the system's geometry. The fact that the optical He II emission line is also affected by the X-ray emission state suggests that the \ion{He}{ii} line actually originates in the stellar wind or in the accretion stream. Radiation from the compact object can decrease the wind velocity; then, due to the interaction between the accretion disk and the wind, changes in luminosity and spectral state may both occur, implying a possible optical - X-ray correlation. This effect has been documented for Cygnus X-3 by \citet{vilhu2021wind}. This phenomenon does not affect the RV measurements, not even within the same phase---the system can be in different emission states, and that would lead to different values of EW. The X-ray and optical data are certainly not contemporaneous, hence more simultaneous observation in X-ray/optical wavelengths are needed for IC 10 X-1 and other WR+BH HMXB systems to support this result.

The stellar winds of the mass donor stars in HMXBs can be highly ionized by the X-rays emanating from the compact objects \citep{HatchettMcCray1977} and resulting in a variety of effects, which range from disruption of the line driving mechanism to the launching of X-ray driven winds and flows within the binary. Only in the part of the X-ray shadow of the WR star will the wind remain largely unaffected \citep[see Figure 1 in][]{laycock2015revisiting}. It has been proposed that the recombination of ions will occur in this shielded (cooler) sector \citep{van1992infrared, van1993spectroscopic, van1996wolf}, resulting in line emission such as He II. The wavelength and shape of the \ion{He}{ii} line will have contributions from the wind velocity, the orbital motion, and the stellar rotation. In such a case, the wind velocity is substantially greater than the orbital motion and the stellar rotation; thus, the observed RV curve will trace the wind velocity vector in projection as the binary rotates  \citep{laycock2015revisiting}. As a result, i) when the WR star lies between the observer and the compact object (mid X-ray eclipse, $\Phi=0.5$), the wind velocity vector will lie along our line of sight producing maximum blueshift; ii) at quadratures ($\Phi = 0.25$ or 0.75), the wind in the shadowed part (and the wind velocity vector) will be orthogonal to the observer's line of sight, hence, producing a zero Doppler shift; and iii) at inferior conjunction of the WR star (the X-ray maximum, $\Phi = 0$ and 1), the shadowed wind is moving away from the observer, resulting in maximum redshift. This phenomenon explains the 0.25 phase shift between the RV curve and the X-ray eclipse in the binary systems IC 10 X-1 \citep{laycock2015revisiting} and NGC 300 X-1 \citep{Carpano2019} as well. However if the phase shift is not exactly 0.25 then the door opens to other phenomena. Indeed, it would be surprising if the conditions for line emission did not exist in other places in the binary. Multiple emitting structures will show up in RV studies by producing additional velocity components (in the same and/or different ion species), each varying at the common binary period, but with different phase offsets and velocity amplitudes, depending on the locations of the emission regions within the binary system. 

Mass transfer for systems such as IC 10 X-1 and NGC 300 X-1 \citep{binder2021wolf} involves not only the ballistic capture of wind material, but additional  radiative and gravitational interactions between the BH and the WR star companion, otherwise the observed luminosity ($L_X$) is too high to be explained \citep{tutukov2016binary}. In the standard  Bondi-Hoyle-Lyttleton accretion theory \citep{bondi1944mechanism, edgar2004review} the capture rate of wind by the compact object is sensitive to the relative velocity between them. In simple terms, only where the wind is moving slower than the local escape velocity, as it sweeps past the compact object, will the matter be accreted. For the fast line-driven winds of supergiant stars, the relative velocity will be high, limiting ballistic capture of wind material, hence suppressing accretion-powered X-ray luminosity. This effect can be seen in the contrast between supergiant and Be-star fed X-ray pulsars \citep{bozzo2016clumpy,corbet1986three}; the latter having slower winds and higher $L_X$. In order to explain the observed X-ray luminosity of IC 10 X-1 and NGC 300 X-1, in the absence of Roche Lobe overflow, the WR wind must be decelerated somehow, before it reaches the compact object. X-ray irradiation will achieve this, both by exerting radiation pressure and by disrupting the line driving process through ionization, so that a persistent  accretion disk can form around the compact object.

The gravitational effects of the compact object can strongly focus the slow moving wind towards the accretor. This streaming material forms a downstream wake, where it stagnates, before flowing into  an accretion stream that feeds the BH accretion disk. Due to the orbital motion of the system, the compact object has already moved when the material impacts the accretion disk, so any hotspot that forms at the impact site is not facing the donor star. This phenomenon has been modeled for supergiant XRBs like Vela X-1 and Cygnus X-1 \citep{el2019formation, huarte2013formation}.
Somewhat counter-intuitively, the direction of accretion flow points {\it inward} with respect to the binary, having passed the BH, stagnated, and reversed direction. 

The accretion stream can be physically separated to a large extent from the BH accretion disk, so there is a possibility of phase lag between the BH, hotspot, and the accretion stream. If \ion{He}{ii} line emission originates in this lagging portion of the accretion stream, then the observed RV curve will also have a phase lag relative to the orbital motion of the BH.

\cite{binder2021wolf} identified and traced the RV curves of three distinct  emission lines in  NGC 300 X-1, using HST to obtain phase-resolved spectra in the near ultraviolet, and published RV values for the visible He II $\lambda$4686 line. The UV spectra revealed the  He II $\lambda$1640 line moving in sync with its visible $\lambda$4686 counterpart, at the now familiar phase shift placing maximum blueshift {\it close to} mid X-ray eclipse, suggesting a common location for visible and ultraviolet He II emission. The third line found by \cite{binder2011deep} was the C IV $\lambda$1550 line, the phase curve of which placed its probable origin on the WR star, under the assumption, that the emission is azimuthally symmetric. Several other weak emission lines were identified in the UV spectra, which could not be traced through a full orbit. 

\cite{binder2021wolf} also noted that the C IV and He II RV curves were offset by 0.3 in phase (rather than exactly 0.25) and they chose to interpret the NGC 300 X-1 data  in the context of a  wind-capture model for accretion in HMXBs \citep{el2019formation, el2019wind, huarte2013formation}, instead of the simpler irradiation and shadowing model discussed by \cite{van1996wolf} and \cite{laycock2015revisiting}. In the wind capture paradigm, the He II fluorescence lines originate in a flow close to the hotspot, where the mass transfer stream coming from the WR star impacts the BH's outer accretion disk.
A schematic for NGC 300 X-1 based upon the above works was presented by \cite{binder2021wolf} (their figure 16). During phases around 0.25 the hotspot is facing us (the observer), and during phases around 0.75 the hotspot lies on the opposite side of the accretion disk. During X-ray eclipse itself, the accretion disk, along with its hotspot, is hidden from direct view; but  emission is visible nonetheless, partly because of the significant vertical extent in the accretion stream, and partly because the eclipse is not total. 

At mid X-ray eclipse ($\Phi=0.5$) the hotspot is approaching inferior conjunction, yet the BH is already there with RV$_{\rm BH}=0$. At this same instant, the hotspot is on the trailing side of the accretion disk as seen by the observer. Therefore the hotspot still has some motion away from us because it has not reached inferior conjunction, so it will show a small positive velocity (RV$_{\rm hotspot}$). On the other hand, if the emission arises in matter streaming back toward the hotspot, then the observed velocity is reversed, and a blueshift is observed.  

A key supporting argument in the model presented by \cite{binder2021wolf} is that the He II line strengths should vary with phase due to obscuration by the accretion disk. When the hotspot is on the side of the disk facing us (leading up to phase 0.25 and for some time afterwards), strong lines are seen; whereas when the hotspot is on the far side of the disk (some duration on either side of phase 0.75), markedly weaker lines (and larger error bars) are observed.  

 A realistic scenario should involve a combination of features from both models summarized above. That the phase shift in IC 10 X-1  matches the wind-shadow model's prediction seems too much of a coincidence, while the presence of an accretion disk also appears necessary on $L_X$ grounds. The parameters of the accretion-stream model meanwhile must be fine-tuned to produce the observed phase shift. 
 
 One can look at what is known about the velocity profiles of WR  stellar winds in more detail and compare them to the spectra. The wind velocity of a WR star at a distance $r$ is given by $v_{\rm w}(r) = v_0 + (v_\infty - v_0)(1 - \frac{R}{r})^\beta$, where $v_\infty$ is the terminal velocity, $R$ is the stellar radius, the preferred value of $\beta=$ 0.8 for these systems, and $v_0$ is the wind velocity at the surface of the star \citep{lamersBook}. \citet{carpano200733} considered $v_0 \sim 0.01 v_\infty$ \textbf{for their study on NGC 300 X-1, and \citet{clark2004wolf} derived $v_\infty = 1750$ km s$^{-1}$ for IC 10 X-1}. There can be a trailing wake formed in the stellar wind \citep{blondin1990hydrodynamic,kaper1994spectroscopic} and another one near the hotspot of the compact object\citep{binder2021wolf,el2019wind}, and the emission line can be a resultant of both regions. The observed velocity can be a vector sum of multiple velocity components arising from different regions of the system. Hence the emission line gets skewed in such a way (due to these multiple components) that results in the observed radial velocity curve. We will return to this issue of overlapping emitters in another publication.

\section{Conclusions}


The \ion{He}{ii}\,$\lambda$4686 line of IC 10 X-1 has been analyzed in detail using 52 GMOS archival spectra. The spectral features are an important probe toward understanding the stellar wind and accretion flow structures. The radial velocity (RV) curve extracted from the \ion{He}{ii}\,$\lambda$4686 line shows a similar 0.25 phase shift as was also seen in the \citet{silverman2008} data, when they were folded using the ephemeris of \citet{laycock2015}. Hence, this behavior is consistent over nearly two-decades-long data, showing a consistency in the system's behavior over time.

We found that the line skewness measurements are approximately anti-correlated (Figure \ref{fig:skew_phase}) with the RV values, in-between the quadratures and through the mid-eclipse at phase 0.5; and correlated for the rest of the orbit. Skewness reaches its maximum positive value at or shortly after phase 0.5 (Figure \ref{fig:skew_phase}), and a most negative value at phases 0.25 and 0.75. The period of the skewness variation with phase is nearly half of that of the RV variation. We consider that the varying  profile asymmetry implied by this result is due to a second emission-line component originating in a different place in the binary, blended with the stronger component that comes from the wind shadow. In this scenario, the positive skewness at max-blueshift of the wind line and negative skewness at max-red-shift of the wind line indicate that the second component moves in anti-phase with the wind line. The faster variation of the skewness can be explained because of the similar appearance of the two emitters, which mimics the situation of a single emitter crossing the observer's line of sight twice within one single orbit. A second emitter on the trailing edge of the BH accretion disk could plausibly generate the observed pattern in skewness.   

From a purely qualitative viewpoint, the same relationship could also be generated if the second emitter is simply static, so that the wind line moves back-and-forth across the static component in velocity space.



In order to fully  understand these observations, one needs to model the combined effect of ionization on the WR wind and emission from the accretion hotspot. We urge observers to obtain optical spectra at the crucial $\Phi = 0.5$ mid-eclipse phase bin, which has not yet been observed in IC 10 X-1. The line profile at that point would provide the cleanest look at the wind in the X-ray shadow region. To systematically map the spatial and velocity distribution of matter in the orbital plane requires additional spectra to fill the phase bins (see Figure  \ref{fig:ic10_hist}), which would also enable Doppler tomography.

Interestingly, we found that the EW values vary with the state of the system's X-ray emission. The harder the X-ray spectrum, the more negative the EW. Yet, this change in EW does not affect the measurements of skewness, which means that for the \ion{He}{ii}\,$\lambda$4686 line, only the total flux is affected by the X-ray state, and not the profile variability, which continues to trace the wind interaction region.

Our results confirm prior assumptions about the WR wind's effects, and open up new avenues toward a better understanding of this system and other WR+BH HMXBs as well. In order to obtain more emission lines for a better mass determination, we need additional multiwavelength observations from \textit{GEMINI}, \textit{Hubble}, and, \textit{JWST}.

\section*{Acknowledgements}
We thank the anonymous reviewer for comments and
suggestions that improved this manuscript. We thank UMass Lowell and the Lowell Center for Space Sciences and Technology for supporting this research. This work was also supported in part by NSF-AAG grant 2109004.

This work was also supported by the international \textit{GEMINI} Observatory, a program of NSF’s NOIRLab, which is managed by the Association of Universities for Research in Astronomy (AURA) under a cooperative agreement with the National Science Foundation, on behalf of the \textit{GEMINI} partnership of Argentina, Brazil, Canada, Chile, the Republic of Korea, and the United States of America. This work was enabled by observations made from the Gemini North telescope, located within the Maunakea Science Reserve and adjacent to the summit of Maunakea. 
\section*{Software and third party data repository citations}

This work made use of Astropy ({\tt www.astropy.org}) a community-developed core Python package and an ecosystem of tools and resources for astronomy \citep{astropy:2013, astropy:2018, astropy:2022}. Data were obtained from the Chandra Data Archive, and software were provided by the Chandra X-ray Center (CXC) in the application packages CIAO \citep{ciao2006}.

The raw data were downloaded from the {\it GEMINI} telescope data archive. The extracted spectra are available in this repository \citep{spec_data} in fits format, other processed data and products (extracted spectra in ascii, or pdf formats, tables and figures) can be acquired by contacting the corresponding author.





\bibliography{reference}{}

\begin{thebibliography}{}
\expandafter\ifx\csname natexlab\endcsname\relax\def\natexlab#1{#1}\fi
\providecommand{\url}[1]{\href{#1}{#1}}
\providecommand{\dodoi}[1]{doi:~\href{http://doi.org/#1}{\nolinkurl{#1}}}
\providecommand{\doeprint}[1]{\href{http://ascl.net/#1}{\nolinkurl{http://ascl.net/#1}}}
\providecommand{\doarXiv}[1]{\href{https://arxiv.org/abs/#1}{\nolinkurl{https://arxiv.org/abs/#1}}}

\bibitem[{{Astropy Collaboration} {et~al.}(2013){Astropy Collaboration},
  {Robitaille}, {Tollerud}, {Greenfield}, {Droettboom}, {Bray}, {Aldcroft},
  {Davis}, {Ginsburg}, {Price-Whelan}, {Kerzendorf}, {Conley}, {Crighton},
  {Barbary}, {Muna}, {Ferguson}, {Grollier}, {Parikh}, {Nair}, {Unther},
  {Deil}, {Woillez}, {Conseil}, {Kramer}, {Turner}, {Singer}, {Fox}, {Weaver},
  {Zabalza}, {Edwards}, {Azalee Bostroem}, {Burke}, {Casey}, {Crawford},
  {Dencheva}, {Ely}, {Jenness}, {Labrie}, {Lim}, {Pierfederici}, {Pontzen},
  {Ptak}, {Refsdal}, {Servillat}, \& {Streicher}}]{astropy:2013}
{Astropy Collaboration}, {Robitaille}, T.~P., {Tollerud}, E.~J., {et~al.} 2013,
  A\&A, 558, A33, \dodoi{10.1051/0004-6361/201322068}

\bibitem[{{Astropy Collaboration} {et~al.}(2018){Astropy Collaboration},
  {Price-Whelan}, {Sip{\H{o}}cz}, {G{\"u}nther}, {Lim}, {Crawford}, {Conseil},
  {Shupe}, {Craig}, {Dencheva}, {Ginsburg}, {Vand erPlas}, {Bradley},
  {P{\'e}rez-Su{\'a}rez}, {de Val-Borro}, {Aldcroft}, {Cruz}, {Robitaille},
  {Tollerud}, {Ardelean}, {Babej}, {Bach}, {Bachetti}, {Bakanov}, {Bamford},
  {Barentsen}, {Barmby}, {Baumbach}, {Berry}, {Biscani}, {Boquien}, {Bostroem},
  {Bouma}, {Brammer}, {Bray}, {Breytenbach}, {Buddelmeijer}, {Burke},
  {Calderone}, {Cano Rodr{\'\i}guez}, {Cara}, {Cardoso}, {Cheedella}, {Copin},
  {Corrales}, {Crichton}, {D'Avella}, {Deil}, {Depagne}, {Dietrich}, {Donath},
  {Droettboom}, {Earl}, {Erben}, {Fabbro}, {Ferreira}, {Finethy}, {Fox},
  {Garrison}, {Gibbons}, {Goldstein}, {Gommers}, {Greco}, {Greenfield},
  {Groener}, {Grollier}, {Hagen}, {Hirst}, {Homeier}, {Horton}, {Hosseinzadeh},
  {Hu}, {Hunkeler}, {Ivezi{\'c}}, {Jain}, {Jenness}, {Kanarek}, {Kendrew},
  {Kern}, {Kerzendorf}, {Khvalko}, {King}, {Kirkby}, {Kulkarni}, {Kumar},
  {Lee}, {Lenz}, {Littlefair}, {Ma}, {Macleod}, {Mastropietro}, {McCully},
  {Montagnac}, {Morris}, {Mueller}, {Mumford}, {Muna}, {Murphy}, {Nelson},
  {Nguyen}, {Ninan}, {N{\"o}the}, {Ogaz}, {Oh}, {Parejko}, {Parley}, {Pascual},
  {Patil}, {Patil}, {Plunkett}, {Prochaska}, {Rastogi}, {Reddy Janga},
  {Sabater}, {Sakurikar}, {Seifert}, {Sherbert}, {Sherwood-Taylor}, {Shih},
  {Sick}, {Silbiger}, {Singanamalla}, {Singer}, {Sladen}, {Sooley},
  {Sornarajah}, {Streicher}, {Teuben}, {Thomas}, {Tremblay}, {Turner},
  {Terr{\'o}n}, {van Kerkwijk}, {de la Vega}, {Watkins}, {Weaver}, {Whitmore},
  {Woillez}, {Zabalza}, \& {Astropy Contributors}}]{astropy:2018}
{Astropy Collaboration}, {Price-Whelan}, A.~M., {Sip{\H{o}}cz}, B.~M., {et~al.}
  2018, AJ, 156, 123, \dodoi{10.3847/1538-3881/aabc4f}

\bibitem[{{Astropy Collaboration} {et~al.}(2022){Astropy Collaboration},
  {Price-Whelan}, {Lim}, {Earl}, {Starkman}, {Bradley}, {Shupe}, {Patil},
  {Corrales}, {Brasseur}, {N{"o}the}, {Donath}, {Tollerud}, {Morris},
  {Ginsburg}, {Vaher}, {Weaver}, {Tocknell}, {Jamieson}, {van Kerkwijk},
  {Robitaille}, {Merry}, {Bachetti}, {G{"u}nther}, {Aldcroft},
  {Alvarado-Montes}, {Archibald}, {B{'o}di}, {Bapat}, {Barentsen}, {Baz{'a}n},
  {Biswas}, {Boquien}, {Burke}, {Cara}, {Cara}, {Conroy}, {Conseil}, {Craig},
  {Cross}, {Cruz}, {D'Eugenio}, {Dencheva}, {Devillepoix}, {Dietrich},
  {Eigenbrot}, {Erben}, {Ferreira}, {Foreman-Mackey}, {Fox}, {Freij}, {Garg},
  {Geda}, {Glattly}, {Gondhalekar}, {Gordon}, {Grant}, {Greenfield}, {Groener},
  {Guest}, {Gurovich}, {Handberg}, {Hart}, {Hatfield-Dodds}, {Homeier},
  {Hosseinzadeh}, {Jenness}, {Jones}, {Joseph}, {Kalmbach}, {Karamehmetoglu},
  {Ka{l}uszy{'n}ski}, {Kelley}, {Kern}, {Kerzendorf}, {Koch}, {Kulumani},
  {Lee}, {Ly}, {Ma}, {MacBride}, {Maljaars}, {Muna}, {Murphy}, {Norman},
  {O'Steen}, {Oman}, {Pacifici}, {Pascual}, {Pascual-Granado}, {Patil},
  {Perren}, {Pickering}, {Rastogi}, {Roulston}, {Ryan}, {Rykoff}, {Sabater},
  {Sakurikar}, {Salgado}, {Sanghi}, {Saunders}, {Savchenko}, {Schwardt},
  {Seifert-Eckert}, {Shih}, {Jain}, {Shukla}, {Sick}, {Simpson},
  {Singanamalla}, {Singer}, {Singhal}, {Sinha}, {Sip{H{o}}cz}, {Spitler},
  {Stansby}, {Streicher}, {{{S}}umak}, {Swinbank}, {Taranu}, {Tewary},
  {Tremblay}, {Val-Borro}, {Van Kooten}, {Vasovi{'c}}, {Verma}, {de Miranda
  Cardoso}, {Williams}, {Wilson}, {Winkel}, {Wood-Vasey}, {Xue}, {Yoachim},
  {Zhang}, {Zonca}, \& {Astropy Project Contributors}}]{astropy:2022}
{Astropy Collaboration}, {Price-Whelan}, A.~M., {Lim}, P.~L., {et~al.} 2022,
  ApJ, 935, 167, \dodoi{10.3847/1538-4357/ac7c74}

\bibitem[{Barnard {et~al.}(2014)Barnard, Steiner, Prestwich, Stevens, Clark, \&
  Kolb}]{barnard2014energy}
Barnard, R., Steiner, J.~F., Prestwich, A.~F., {et~al.} 2014, ApJ, 792, 131

\bibitem[{Bauer \& Brandt(2004)}]{bauer2004chandra}
Bauer, F.~E., \& Brandt, W.~N. 2004, ApJ, 601, L67

\bibitem[{Belczynski {et~al.}(2013)Belczynski, Bulik, Mandel, Sathyaprakash,
  Zdziarski, \& Miko{\l}ajewska}]{belczynski2013cyg}
Belczynski, K., Bulik, T., Mandel, I., {et~al.} 2013, ApJ, 764, 96

\bibitem[{Bhattacharya(2022)}]{spec_data}
Bhattacharya, S. 2022, {Replication Data For ``Probing the Stellar Wind of the
  Wolf-Rayet Star in IC 10 X-1''}, V1,  Harvard Dataverse,
  \dodoi{10.7910/DVN/U3PFZB}

\bibitem[{Bhattacharya {et~al.}(2021)Bhattacharya, Chene, Roy, Cappallo,
  Laycock, \& Christodoulou}]{bhattacharya2021mass}
Bhattacharya, S., Chene, A.-N., Roy, A., {et~al.} 2021, in APS April Meeting
  Abstracts, Vol. 2021, H09--008

\bibitem[{Bhattacharya {et~al.}(2022)Bhattacharya, Laycock, Chene, Binder, \&
  Christodoulou}]{bhattacharya2022stellar}
Bhattacharya, S., Laycock, S.~G., Chene, A.-N., Binder, B., \& Christodoulou,
  D. 2022, Bulletin of the American Physical Society

\bibitem[{Binder {et~al.}(2011)Binder, Williams, Eracleous, Garcia, Anderson,
  \& Gaetz}]{binder2011deep}
Binder, B.~A., Williams, B.~F., Eracleous, M., {et~al.} 2011, ApJ, 742, 128

\bibitem[{Binder {et~al.}(2021)Binder, Sy, Eracleous, Christodoulou,
  Bhattacharya, Cappallo, Laycock, Plucinsky, \& Williams}]{binder2021wolf}
Binder, B.~A., Sy, J.~M., Eracleous, M., {et~al.} 2021, ApJ, 910, 74

\bibitem[{Blondin {et~al.}(1990)Blondin, Kallman, Fryxell, \&
  Taam}]{blondin1990hydrodynamic}
Blondin, J.~M., Kallman, T.~R., Fryxell, B.~A., \& Taam, R.~E. 1990, ApJ, 356,
  591

\bibitem[{Bondi \& Hoyle(1944)}]{bondi1944mechanism}
Bondi, H., \& Hoyle, F. 1944, MNRAS, 104, 273

\bibitem[{Bozzo {et~al.}(2016)Bozzo, Oskinova, Feldmeier, \&
  Falanga}]{bozzo2016clumpy}
Bozzo, E., Oskinova, L., Feldmeier, A., \& Falanga, M. 2016, A\&A, 589, A102

\bibitem[{Brandt {et~al.}(1997)Brandt, Ward, Fabian, \&
  Hodge}]{brandt1997rosat}
Brandt, W.~N., Ward, M.~J., Fabian, A.~C., \& Hodge, P.~W. 1997, MNRAS, 291,
  709

\bibitem[{Brown {et~al.}(2000)Brown, Lee, Wijers, Lee, Israelian, \&
  Bethe}]{brown2000grb}
Brown, G.~E., Lee, C.~H., Wijers, R.~A., {et~al.} 2000, NewA, 5, 191

\bibitem[{{Carpano} {et~al.}(2019){Carpano}, {Haberl}, {Crowther}, \&
  {Pollock}}]{Carpano2019}
{Carpano}, S., {Haberl}, F., {Crowther}, P., \& {Pollock}, A. 2019, IAU
  Symposium, 346, 187, \dodoi{10.1017/S1743921318007615}

\bibitem[{Carpano {et~al.}(2007)Carpano, Pollock, Prestwich, Crowther, Wilms,
  Yungelson, \& Ehle}]{carpano200733}
Carpano, S., Pollock, A.~M., Prestwich, A., {et~al.} 2007, A\&A, 466, L17

\bibitem[{Clark \& Crowther(2004)}]{clark2004wolf}
Clark, J.~S., \& Crowther, P.~A. 2004, A\&A, 414, L45

\bibitem[{Corbet(1986)}]{corbet1986three}
Corbet, R.~H. 1986, MNRAS, 220, 1047

\bibitem[{Crowther {et~al.}(2003)Crowther, Drissen, Abbott, Royer, \&
  Smartt}]{crowther2003gemini}
Crowther, P.~A., Drissen, L., Abbott, J.~B., Royer, P., \& Smartt, S.~J. 2003,
  A\&A, 404, 483

\bibitem[{Edgar(2004)}]{edgar2004review}
Edgar, R. 2004, NewAR, 48, 843

\bibitem[{El~Mellah {et~al.}(2019{\natexlab{a}})El~Mellah, Sander, Sundqvist,
  \& Keppens}]{el2019formation}
El~Mellah, I., Sander, A. A.~C., Sundqvist, J.~O., \& Keppens, R.
  2019{\natexlab{a}}, A\&A, 622, A189

\bibitem[{El~Mellah {et~al.}(2019{\natexlab{b}})El~Mellah, Sundqvist, \&
  Keppens}]{el2019wind}
El~Mellah, I., Sundqvist, J.~O., \& Keppens, R. 2019{\natexlab{b}}, A\&A, 622,
  L3

\bibitem[{{Fruscione} {et~al.}(2006){Fruscione}, {McDowell}, {Allen},
  {Brickhouse}, {Burke}, {Davis}, {Durham}, {Elvis}, {Galle}, {Harris},
  {Huenemoerder}, {Houck}, {Ishibashi}, {Karovska}, {Nicastro}, {Noble},
  {Nowak}, {Primini}, {Siemiginowska}, {Smith}, \& {Wise}}]{ciao2006}
{Fruscione}, A., {McDowell}, J.~C., {Allen}, G.~E., {et~al.} 2006, in Society
  of Photo-Optical Instrumentation Engineers (SPIE) Conference Series, Vol.
  6270, Society of Photo-Optical Instrumentation Engineers (SPIE) Conference
  Series, ed. D.~R. {Silva} \& R.~E. {Doxsey}, 62701V,
  \dodoi{10.1117/12.671760}

\bibitem[{{Hatchett} \& {McCray}(1977)}]{HatchettMcCray1977}
{Hatchett}, S., \& {McCray}, R. 1977, ApJ, 211, 552, \dodoi{10.1086/154962}

\bibitem[{Huarte-Espinosa {et~al.}(2013)Huarte-Espinosa, Carroll-Nellenback,
  Nordhaus, Frank, \& Blackman}]{huarte2013formation}
Huarte-Espinosa, M., Carroll-Nellenback, J., Nordhaus, J., Frank, A., \&
  Blackman, E.~G. 2013, MNRAS, 433, 295

\bibitem[{Iben~Jr \& Tutukov(1984)}]{iben1984supernovae}
Iben~Jr, I., \& Tutukov, A.~V. 1984, ApJS, 54, 335

\bibitem[{Igoshev {et~al.}(2021)Igoshev, Chruslinska, Dorozsmai, \&
  Toonen}]{igoshev2021combined}
Igoshev, A.~P., Chruslinska, M., Dorozsmai, A., \& Toonen, S. 2021, MNRAS, 508,
  3345

\bibitem[{Israelian {et~al.}(1999)Israelian, Rebolo, Basri, Casares, \&
  Martin}]{israelian1999}
Israelian, G., Rebolo, R., Basri, G., Casares, J., \& Martin, E.~L. 1999,
  Nature, 401, 142

\bibitem[{Kaper {et~al.}(1994)Kaper, Hammerschlag-Hensberge, \&
  Zuiderwijk}]{kaper1994spectroscopic}
Kaper, L., Hammerschlag-Hensberge, G.~C., \& Zuiderwijk, E.~J. 1994, A\&A, 289,
  846

\bibitem[{Koljonen \& Maccarone(2017)}]{koljonen2017gemini}
Koljonen, K.~I., \& Maccarone, T.~J. 2017, MNRAS, 472, 2181

\bibitem[{{Lamers} \& {Cassinelli}(1999)}]{lamersBook}
{Lamers}, H. J.~G.~L.~M., \& {Cassinelli}, J.~P. 1999, {Introduction to Stellar
  Winds}

\bibitem[{Laycock {et~al.}(2014)Laycock, Cappallo, Oram, \&
  Balchunas}]{laycock2014transient}
Laycock, S., Cappallo, R., Oram, K., \& Balchunas, A. 2014, The Astrophysical
  Journal, 789, 64

\bibitem[{Laycock {et~al.}(2015{\natexlab{a}})Laycock, Cappallo, \&
  Moro}]{laycock2015}
Laycock, S.~G., Cappallo, R.~C., \& Moro, M.~J. 2015{\natexlab{a}}, MNRAS, 446,
  1399

\bibitem[{Laycock {et~al.}(2017)Laycock, Cappallo, Williams, Prestwich, Binder,
  \& Christodoulou}]{laycock2017x}
Laycock, S.~G., Cappallo, R.~C., Williams, B.~F., {et~al.} 2017, ApJ, 836, 50

\bibitem[{Laycock {et~al.}(2015{\natexlab{b}})Laycock, Maccarone, \&
  Christodoulou}]{laycock2015revisiting}
Laycock, S.~G., Maccarone, T.~J., \& Christodoulou, D.~M. 2015{\natexlab{b}},
  MNRAS Letters, 452, L31

\bibitem[{Orosz {et~al.}(2001)Orosz, Kuulkers, van~der Klis, McClintock,
  Garcia, Callanan, Bailyn, Jain, \& Remillard}]{orosz2001}
Orosz, J.~A., Kuulkers, E., van~der Klis, M., {et~al.} 2001, ApJ, 555, 489

\bibitem[{Park {et~al.}(2006)Park, Kashyap, Siemiginowska, Van~Dyk, Zezas,
  Heinke, \& Wargelin}]{park2006bayesian}
Park, T., Kashyap, V.~L., Siemiginowska, A., {et~al.} 2006, ApJ, 652, 610

\bibitem[{Prestwich {et~al.}(2007)Prestwich, Kilgard, Crowther, Carpano,
  Pollock, Zezas, Saar, Roberts, \& Ward}]{prestwich2007orbital}
Prestwich, A.~H., Kilgard, R., Crowther, P.~A., {et~al.} 2007, ApJL, 669, L21

\bibitem[{{Psaltis}(2006)}]{psaltis}
{Psaltis}, D. 2006, {Accreting neutron stars and black holes: a decade of
  discoveries}, Vol.~39, 1--38

\bibitem[{Remillard \& McClintock(2006)}]{remillard2006x}
Remillard, R.~A., \& McClintock, J.~E. 2006, ARA\&A, 44, 49

\bibitem[{Silverman \& Filippenko(2008)}]{silverman2008}
Silverman, J.~M., \& Filippenko, A.~V. 2008, ApJL, 678, L17

\bibitem[{{Tanaka} \& {Lewin}(1995)}]{tanaka_lewin}
{Tanaka}, Y., \& {Lewin}, W.~H. 1995, in X-ray Binaries, 126--174

\bibitem[{{Tarter} {et~al.}(1969){Tarter}, {Tucker}, \&
  {Salpeter}}]{tarter1969ion}
{Tarter}, C.~B., {Tucker}, W.~H., \& {Salpeter}, E.~E. 1969, Apj, 156, 943,
  \dodoi{10.1086/150026}

\bibitem[{Tauris {et~al.}(2017)Tauris, Kramer, Freire, Wex, Janka, Langer,
  Podsiadlowski, Bozzo, Chaty, Kruckow, {et~al.}}]{tauris2017formation}
Tauris, T.~M., Kramer, M., Freire, P.~C., {et~al.} 2017, ApJ, 846, 170

\bibitem[{Tutukov \& Fedorova(2016)}]{tutukov2016binary}
Tutukov, A.~V., \& Fedorova, A.~V. 2016, ARep, 60, 106

\bibitem[{Van~Kerkwijk(1993)}]{van1993spectroscopic}
Van~Kerkwijk, M.~H. 1993, A\&A, 276, L9

\bibitem[{van Kerkwijk {et~al.}(1996)van Kerkwijk, Geballe, King, van~der Klis,
  \& van Paradijs}]{van1996wolf}
van Kerkwijk, M.~H., Geballe, T.~R., King, D.~L., van~der Klis, M., \& van
  Paradijs, J. 1996, A\&A, 314, 521

\bibitem[{Van~Kerkwijk {et~al.}(1992)Van~Kerkwijk, Charles, Geballe, King,
  Miley, Molnar, Van~den Heuvel, vander Klis, \&
  Van~Paradijs}]{van1992infrared}
Van~Kerkwijk, M.~H., Charles, P.~A., Geballe, T.~R., {et~al.} 1992, Nature,
  355, 703

\bibitem[{Vilhu {et~al.}(2021)Vilhu, Kallman, Koljonen, \&
  Hannikainen}]{vilhu2021wind}
Vilhu, O., Kallman, T.~R., Koljonen, K.~I., \& Hannikainen, D.~C. 2021, A\&A,
  649, A176

\bibitem[{Whelan \& Iben~Jr(1973)}]{whelan1973binaries}
Whelan, J., \& Iben~Jr, I. 1973, ApJ, 186, 1007

\end{thebibliography}
\bibliographystyle{aasjournal}



\end{document}